\documentclass[final,5p,times,twocolumn]{elsarticle}

\usepackage{amssymb}

\begin{document}

\begin{frontmatter}



\title{ Solar neutrino results from Borexino and main future perspectives }


\author{ Marco Pallavicini }

\address{  Dipartimento di Fisica,
Universit\`a di Genova e INFN
\\
 Genova, via Dodecaneso, 33 - I-16146, Italy\\
 {\rm E-mail: marco.pallavicini@ge.infn.it \\}
 { ~ \\}
 {\large on behalf of the Borexino Collaboration} }
 
\author[mi]{G. Bellini}
\author[pri1]{J. Benziger} 
\author[mi]{ S. Bonetti} 
\author[mi]{ M. Buizza Avanzini} 
\author[mi]{ B. Caccianiga} 
\author[mass]{ L. Cadonati} 
\author[pri2]{ F. Calaprice} 
\author[ge]{ C. Carraro} 
\author[pri2]{ A. Chavarria} 
\author[pri2]{ F. Dalnoki-Veress} 
\author[mi]{ D. D{\textquoteright}Angelo} 
\author[apc]{ H. de Kerret} 
\author[ge]{ S. Davini} 
\author[du]{ A. Derbin} 
\author[ku]{ A. Etenko} 
\author[du]{ K. Fomenko} 
\author[mi]{ D. Franco} 
\author[pri2]{ C. Galbiati} 
\author[lngs]{ S. Gazzana} 
\author[mi]{M. Giammarchi} 
\author[mo]{ M. Goeger-Neff} 
\author[pri2]{ A. Goretti} 
\author[vt]{ C. Grieb} 
\author[ge]{ E. Guardincerri} 
\author[vt]{ S. Hardy} 
\author[lngsi]{ Aldo Ianni} 
\author[pri2]{Andrea Ianni} 
\author[vt]{ M. Joyce} 
\author[mi]{ V. Kobychev} 
\author[lngs]{ G. Korga} 
\author[apc]{ D. Kryn} 
\author[lngs]{ M. Laubenstein} 
\author[pri2]{ M. Leung} 
\author[mo]{ T. Lewke} 
\author[ku]{ E. Litvinovich} 
\author[pri2]{ B. Loer} 
\author[mi]{ P. Lombardi} 
\author[mi]{ L. Ludhova} 
\author[ku]{ I. Machulin} 
\author[vt]{ S. Manecki} 
\author[he]{ W. Maneschg} 
\author[ge]{ G. Manuzio} 
\author[pg]{ F. Masetti} 
\author[pri2]{ K. McCarty} 
\author[mo]{ Q. Meindl}
\author[mi]{ E. Meroni} 
\author[mi]{ L. Miramonti} 
\author[cra]{ M. Misiaszek} 
\author[lngs]{ D. Montanari} 
\author[du]{ V. Muratova} 
\author[mo]{ L.Oberauer} 
\author[apc]{ M. Obolensky} 
\author[pg]{ F. Ortica} 
\author[ge]{ M. Pallavicini} 
\author[lngs]{ L. Papp} 
\author[mi]{ L. Perasso} 
\author[ge]{ S. Perasso} 
\author[pri2]{ A. Pocar} 
\author[vt]{ R.S. Raghavan} 
\author[mi]{ G. Ranucci} 
\author[lngs]{ A. Razeto} 
\author[ge]{ P. Risso} 
\author[pg]{ A. Romani} 
\author[vt]{ D. Rountree} 
\author[ku]{ A. Sabelnikov} 
\author[pri2]{ R. Saldanha} 
\author[ge]{ C. Salvo} 
\author[he]{ S. Schonert} 
\author[he]{ H. Simgen} 
\author[ku]{ M. Skorokhvatov} 
\author[du]{ O. Smirnov} 
\author[du]{ A. Sotnikov} 
\author[ku]{ S. Sukhotin} 
\author[lngs]{ Y. Suvorov} 
\author[lngs]{ R. Tartaglia} 
\author[ge]{G. Testera}
\author[apc]{ D. Vignaud} 
\author[vt]{ R.B. Vogelaar} 
\author[mo]{ F. von Feilitzsch} 
\author[cra]{ M. Wojcik} 
\author[mo]{ M. Wurm} 
\author[du]{ O. Zaimidoroga} 
\author[ge]{ S. Zavatarelli} 
\author[he]{G. Zuzel}

\address[lngs]{I.N.F.N. Laboratori Nazionali del Gran Sasso -- Assergi -- Italy}
\address[mi]{Dipartimento di Fisica dell'Universit\`a degli Studi e INFN -- Milano -- Italy}
\address[pri1]{Princeton University, Chemical Engineering Department -- Princeton, NJ -- USA}
\address[pri2]{Princeton University, Physics Department -- Princeton, NJ -- USA}
\address[ku]{RRC Kurchatov Institute -- Moscow -- Russia}
\address[apc]{Laboratoire AstroParticule et Cosmologie APC -- Paris -- France}
\address[peters]{St. Petersburg Nuclear Physics Institute -- Gatchina -- Russia}
\address[mit]{Massachussetts Institute of Technology, Department of Physics -- Cambridge, MA -- USA}
\address[du]{Joint Institute for Nuclear Research -- Dubna -- Russia}
\address[mu]{Technische Universit\"at Muenchen -- Garching -- Germany}
\address[vt]{Virginia Tech,  Physics Department -- Blacksburg, VA -- USA}
\address[ge]{Dipartimento di Fisica dell'Universit\`a e INFN -- Genova -- Italy}
\address[pg]{Dipartimento di Chimica dell'Universit\`a e INFN -- Perugia -- Italy}
\address[cra]{Marian Smoluchowski Institute of Physics, Jagiellonian University -- Krakow -- Poland}
\address[he]{Max-Planck-Institut f\"ur Kernphysik -- Heidelberg -- Germany}
\address[mass]{Department, University of Massachusetts, Amherst, AM01003, USA}

\begin{abstract}
Borexino is a solar neutrino experiment running at the Laboratori Nazionali del Gran Sasso, Italy. 
The radioactive background levels in the liquid scintillator target meet or even exceed design goals, 
opening unanticipated opportunities.
The main results, so far, are the measurement of the $^7$Be solar neutrino flux (the first ever done) and the 
measurement of the $^8$B neutrino flux performed with electron energy threshold of 2.8 MeV.
The short and medium term perspectives are summarized in the conclusions.

\end{abstract}


\end{frontmatter}

\normalsize\baselineskip=15pt

\section{Introduction}

Borexino \cite{BxNim} detects solar neutrinos via their elastic scattering
on the electrons of an ultra-pure liquid scintillator target. 
The main physics goal  is the measurement of the flux and of the energy spectrum of solar 
neutrinos with sub-MeV or few MeV energy. Other goals include geoneutrinos detection, 
super-nova neutrinos detection and the search for very rare decays \cite{CTF1,CTF2,CTF3,CTF4}.

Flavor oscillations of solar neutrinos with MSW effect \cite{MSW} have been well established by radiochemical experiments 
\cite{RadioChemical} and 
by water Cerenkov detectors \cite{Cerenkov}.
The range of the parameters describing the oscillation phenomenon  has been 
constrained by SNO and KamLand\cite{Kamland}
to lie in the so called LMA (Large Mixing Angle)
region of the plane $ \theta_{12}$ $\Delta m^2_{12}$ ($tan^2 (2\theta_{12}) =0.47^{ +0.06}_{-0.05} $ and 
$\Delta m^2_{12} 7.59^{+0.21}_{-0.21} \cdot 10^{-5} eV^2$).

The main prediction of the LMA-MSW are the energy dependence  of the neutrino survival probability $P_{ee}$ and the
lack of day-night asymmetry. The $P_{ee}$ decreases with increasing energy. 
Matter effects dominate at energies above 3 MeV while are absent below 1 MeV. The region in between is called the transition region. While the LMA-MSW predicts a well defined shape for the  $P_{ee}$ function in the transition region,
current experimental data do not constraint it at all, and some theoretical models, 
including non standard interactions, predict survival probability curves  with different shape \cite{NotStandard}.
 
Borexino is presently the only running experiment that measured the signal rate due to the 0.862 MeV  
$^7$Be neutrinos \cite{BxBe}\cite{BxBe2} and the one due to $^8B$ \cite{BxB8} with  a energy threshold (2.8 MeV),
lower than any previous experiment. Besides, a measurement of a null day night difference of the $^7$Be flux 
provides a further new confirmation of the LMA-MSW solution. Interest of this measurement is also related to the 
possibility to accommodate  quite large effects in some alternative oscillation scenario based on the mass varying model
\cite{Holanda}.

The neutrino interaction rate measured by Borexino depends on  the solar neutrino flux and 
on the oscillation parameters.
A scientific debate about high and low metallicity solar models and the related flux 
calculations \cite{carlos1} is in progress. The relevance of the measurements of 
the various components of the solar neutrino measurements in Borexino is then twofold:
from one side they can increase the confidence in the oscillation scenario and from the other side, 
assuming the knowledge of the oscillation parameters, they provide a measurement of the 
absolute solar neutrino flux. The precision of the actual data about solar neutrinos  does not allow 
to distinguish between high and low metallicity models \cite{carlos1} but future high precision 
measurements of the $^7$Be might give useful constraints. 

Besides, the experimental determination of the flux of the CNO components is of 
strategical importance being the CNO flux prediction different by more than $30 \%$ 
in the two classes of models.

 \section{The $^7$Be signal} 
\begin{figure}[hbt!]
\begin{center}
\includegraphics[width=0.52\textwidth]{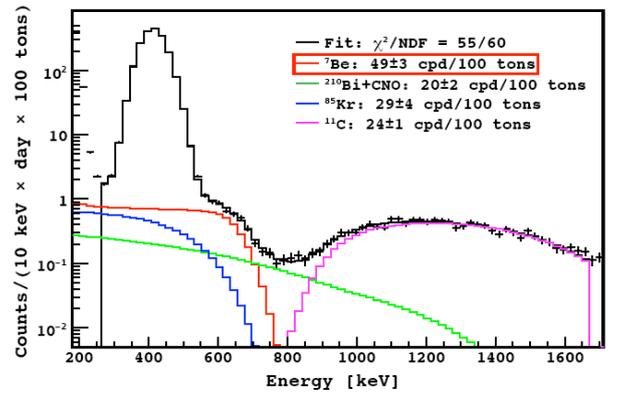}
\caption{The Borexino energy spectrum (192 days of live time) and its fit. A similar fit procedure in which the $^{210}$Po is statistically subtracted by exploiting the $\alpha$-$\beta$ separation capability of the detector gives consistent results.}
\label{Spettro}
\end{center}
\end{figure}

The $^7$Be signal rate in Borexino is obtained by fitting the energy spectrum which is a superposition of the neutrino
signal events and the background. Event selection is described in \cite{BxBe2}. 
The resulting energy spectrum corresponding to a live time of 192 days is shown in figure \ref{Spettro}.

The energy calibration  is obtained by studying the 
 $\beta$ decay of $^{14}C$ with 156 KeV end point (not shown) and through the spectral fit itself.
The expected $^7$Be spectral signature is a electron recoil spectrum with a Compton like shape and its features are  visible in  figure \ref{Spettro}. The large peak in the same figure is due to the 5.3 MeV $\alpha$ decay of $^{210}Po$, 
a daughter of $^{222}Rn$ out of equilibrium.  The ionization quenching of the scintillator reduces the 
visible energy by a factor about 13 and brings the $\alpha$ peak in the energy region of the $^7$Be signal.
A positive side effect of this large background is its use to study the yield stability and the energy resolution 
of the detector. The $^{210}Po$ count rate decreases with time consistently with its mean life of 200 days.

The study of the time correlated events belonging to the $^{238}$U and $^{232}$Th radioactive chains yields, under the hypothesis of secular equilibrium, an internal contamination for $^{238}U$ of $1.6 \pm 0.1 \cdot 10^{-17} g/g$ and for 
$^{232}Th$ of $6.8 \pm 1.5 \cdot 10^{-18} g/g$.
The concentration of these contaminants is more than an order of magnitude lower than 
the design value of $\simeq 10^{-16} g/g$ and it is not therefore the main issue of the $^7$Be analysis.
On the contrary, the most important background  is due to the $\beta$ decay of $^{85}Kr$ with 687 KeV end point
having a rate of the same order of magnitude of the $^7$Be signal and a spectral shape not too different.

The analysis of the rare decays of $^{85}$Kr into $^{85}$Rb (branching ration 0.43 $\%$ but taggable due to the 
presence of time correlated events) yields $28 \pm 7$ counts/(days 100 t) after a live time of more than 1 year. 
Additional background is identified as $^{210}$Bi and as $^{11}$C. The last one is produced by the interaction of 
muons in the scintillator.

The energy spectrum is fitted by two procedures \cite{BxBe}\cite{BxBe2}: one of the them includes the $^{210}$Po $\alpha$ 
peak and in the second one a statistical subtraction of this peak is applied. 
The two results are mutually consistent and they  give the value of the 0.862 $^7$Be solar neutrino interaction rate of 
$49 \pm 3_{stat} \pm 4_{syst} $ ev/(day 100 t) after 192 days of live time. 
Assuming the flux of the Standard Solar Model with high metallicity the expected rate without oscillations is
$74 \pm 4 $ $ev/( day 100 t)$ which should reduce to $48 \pm 4 $ ev/(day 100t)  using the  LMA-MSW oscillation parameters.
The $^7$Be measurement of Borexino  confirms the prediction at low energy of the LMA oscillation model.
  
A big effort is now in progress to reduce the errors associated to the measurement of the $^7$Be signal rate: 
the main contributions are the imperfect knowledge of the fiducial volume and of the detector response function 
(each one of them gives a contribution to the systematic error of $6\%$ ). We are planning to reduce these 
errors through the detector calibration: three calibration campaigns have been already completed, and the results 
are under analysis.
\begin{figure}
\begin{center}
\includegraphics[width = 0.56\textwidth]{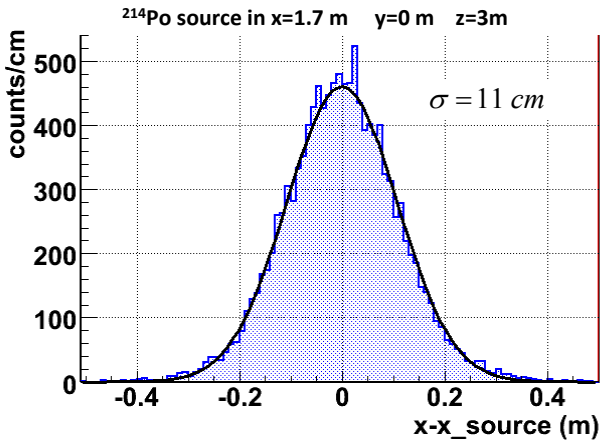}
\caption{As preliminary example of calibration data the distribution of the measured x positions of the scintillation 
events due to a ${214}^Po$ $\alpha$ decay from a Radon source placed off axis is shown. The vertical axis in Borexino 
is z. The x coordinate is reconstructed with a Gaussian distribution having a standard deviation of 11 cm. 
The accuracy of the reconstruction of the absolute position is under study. }
\label{Source}
\end{center}
\end{figure}
\section{Day night asymmetry}
Several different radioactive sources ($^{14}$C source, Rn loaded scintillator source, 
$\gamma$-sources, neutron sources) have been 
inserted in the inner vessel center and along the vertical axis (first calibration campaign)and in more than 
100 positions off axis (second and third calibration campaign) to verify the accuracy of the event position 
reconstruction (see figure \ref{Source}), which is used to define the fiducial volume, to check the absolute 
energy calibration and to study the dependence of the energy response on the scintillation position.
  
Particular care has been devoted to the design of the source insertion system, to the selection of its 
materials and to the definition of the insertion procedure in order  to minimize the risk of introduction of 
any radioactive contaminants in the detector that would spoil the exceptional performances of Borexino.
The post calibration data indicate that   the source insertion operations did not introduce important changes 
of the background. 

Eight $\gamma$-sources with energies ranging from 150 keV up to 2.2 MeV and the
neutron source with its 4.4 MeV capture line on $^{12}$C have been 
chosen in order to study the energy response in the whole region of interest for solar neutrino physics. 

The true source position can be determined within 2 cm accuracy through the use of a red led light (mounted 
on the source support) and a system of CCD camera. The goal is to reduce the error associated to the $^7$Be 
signal rate to few percent.

\begin{figure}
\begin{center}
\rotatebox[origin=rB]{90}{\includegraphics[width = 6 cm]{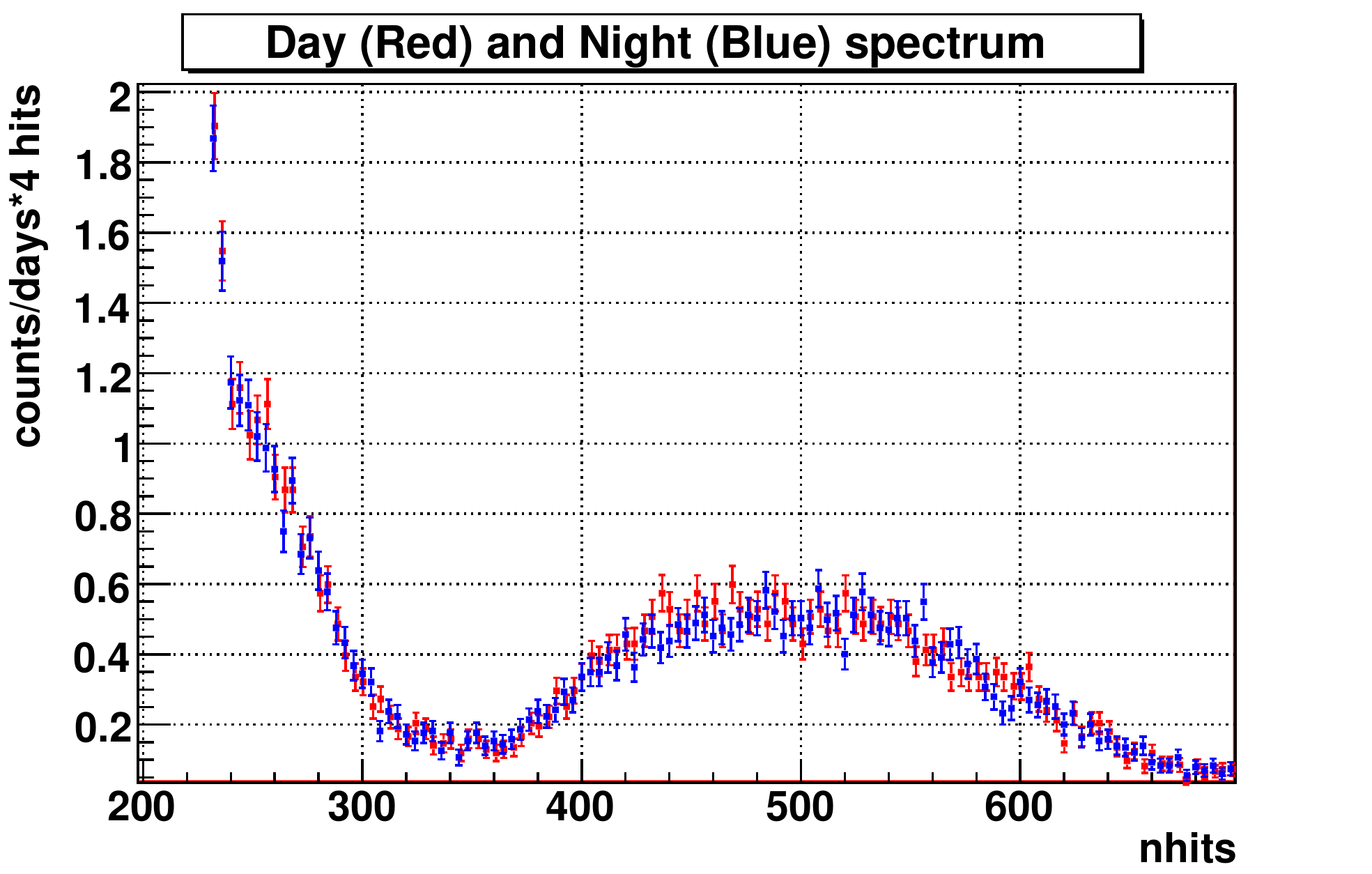}}
\caption{Day (red) and night (blue) spectra of the selected events normalized to the night live time and in the energy region outside the $^{210}$Po peak.}
\label{DNSpectrumLinear}
\end{center}
\end{figure}

\begin{figure}
\begin{center}
\includegraphics[width = 0.5\textwidth]{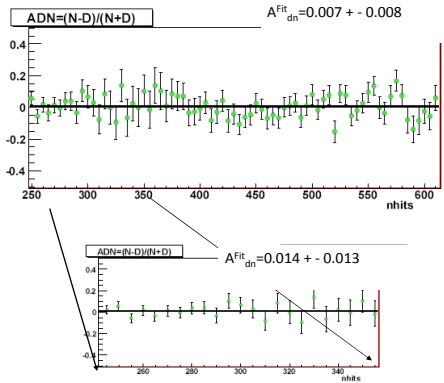}
\caption{Binned day nigh asymmetry as a function of nhits and the zoom in the region where the 
contribution of the $^7$Be signal is maximum. The fit is performed with a constant function. 
The reduced Chi squared is 66.5/72.}
\label{Diff}
\end{center}
\end{figure}
A preliminary analysis of the day and night spectra provides a 
further confirmation of the prediction of the LMA model through the absence of a significant day-night asymmetry in the $^7$Be flux.
For each event,  from the absolute time we compute the  value of the Sun zenith angle at the 
Laboratori Nazionali del G. Sasso latitude.
Figure \ref{DNSpectrumLinear} shows the day and night spectra corresponding to a total live time of 422.12 days with 212.87 days and 209.25 night.
The horizontal axis represents the number of hits detected by the photomultipliers which is closely connected to the energy of the events.
Events are selected as described in \cite{BxBe} the only difference concerns the choice of the fiducial volume being here only the spatial cut r$<$3 m  applied. 
The day-night asymmetry $A_{dn}$ is defined as $A_{dn}=\frac { C_n -C_d} {C_n+C_d} $ where $C_n$ and $C_d$ are
 the counts during day and the night time.
$A_{dn}$ has been evaluated for every bin (see figure \ref{Diff}) in the region from nhits $>$ 250  to nhits $<$ 700 as shown in figure. Here below 
nhits=350  the signal to background ratio is maximum while for  nhits$>$350 the background (due to $^{11}C$) is dominant.
The binned day night asymmetry well fit with a constant function providing $A^{fit}_{dn}=0.007\pm 0.008$ in the region (250,700) nhits and 
$A^{fit}_{dn}=0.014\pm 0.013$ in the region (250,350) nhits. This last value is 
consistent with $A^{aver}_{dn}=0.011\pm 0.014$ obtained using the integrated counts in the region with nhits (250,350).
The absence of significant difference between the day and night signals in the high energy region is a check of the consistency of the data while
the results about $A^{aver}_{dn}$ and $A^{fit}_{dn}$ in the region (250,350) nhits show that the day night asymmetry of the $^7$Be solar 
neutrino signal is zero within one standard deviation.
This result is independent on the precision of the definition of the fiducial volume and on the knowledge of the detector response function.

The day night asymmetry in the region nhits (250,350) here discussed includes  the contribution both of the signal and  of the background. Considering the statistical precision of the $^7$Be flux determination
in the day and night periods we get the contribution of the signal alone  $A^{\nu}_{dn}=0.02 \pm 0.04_{stat}$.
This is our preliminary result confirming the expectation of the LMA-MSW scenario. Further analysis and the evaluation of the contribution of possible systematic
errors due to  the selection of the data sample are in progress.

\section{The low energy threshold $^8$B signal}
The excellent radiopurity levels obtained by Borexino made possible a measurement of the $^8$B solar neutrino 
flux with the unprecedented energy threshold of 2.8 MeV \cite{BxB8}. 
This value is mainly determined by the need to cut the residual $\gamma$ background due to the Thallium decay in 
the PMT materials. The expected signal rate, including neutrino oscillations, is $0.26 \pm 0.03$ counts/day in 100 t.
Data selection procedure (see \cite{BxB8} for details) includes the  removal (with $99.7 \%$ efficiency) 
of short lived ($\tau<2 $ sec cosmogenic isotopes by vetoing the detector for 5 sec after each muon crossing 
the scintillator,  the removal of $^{10}C$ by the triple coincidence with the parent muon and the neutron capture 
on proton and the statistical subtraction of the Thallium spectrum due to the internal radioactivity. 
The resulting $^8B$ neutrinos signal rate  is $0.26 \pm 0.04_{stat} \pm 0.02_{sys}$ counts/(days 100 t) 
after 246 days of live time. This measurement confirms flavor oscillations at 4.2 $\sigma$ level.
 
\begin{figure}[hbt!]
\begin{center}
\includegraphics[width = 0.5\textwidth]{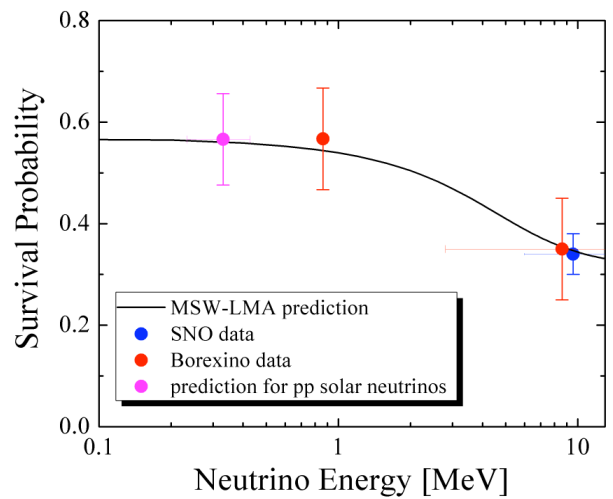}
\caption{The electron neutrino survival probability of the LMA-MSW model and the experimental results including the new data of Borexino}
\label{Survival}
\end{center}
\end{figure}

The neutrino survival probability at an average energy of 8.6 MeV is $P^{^8B}_{ee}=0.35 \pm 0.10$ while
the one of the $^7$Be neutrinos is $P^{^7Be}_{ee}=0.56 \pm 0.10$. 
Eliminating the common sources of systematic errors the ratio between the two probabilities is $1.6 \pm 0.33$ 
confirming the expectation of the LMA-MSW oscillation scenario at $93\%$ C.L. (see figure \ref{Survival}).

\section{Conclusions and future programs }
Borexino has completed the first real time measurements of $^7$Be and the low threshold $^8$B signal rate. 
A preliminary result about the day night asymmetry has also been obtained. All results confirm the current
LMA-MSW scenario. Annual modulation analysis is underway.

A calibration campaign is in progress. The results are expected to contribute to a significant reduction of the 
errors on the fiducial volume and on the detector response function, 
yielding a $^7$Be signal rate measurement with few percent accuracy and of the reduction 
of the systematic errors of the $^8B$ measurement.

The possibility to purify the scintillator to reduce the background due to $^{85}Kr$ and to $^{210}Bi$ is under consideration.
If successful, Borexino might attempt the direct detection of pep and pp, and CNO neutrinos.
  
\section{Acknowledgments}
This work has been funded by: INFN (Italy), NSF (USA), BMBF, DFGandMPG (Germany), Rosnauka(Russia), MNiSW (Poland).

\end{document}